\documentstyle[12pt]{article}
\newcommand{\be}{\begin{equation}}
\newcommand{\ee}{\end{equation}}
\begin{document}
\def\theequation{\arabic{section}.\arabic{equation}}
\begin{titlepage}
\title{Lensing by gravitational waves in scalar--tensor gravity: 
Einstein frame analysis} \author{Valerio Faraoni$^{1}$\\
and\\
Edgard Gunzig$^{2,3}$\\ \\
{\small \it $^1$ Inter-University Centre for Astronomy and Astrophysics}\\
{\small \it Post Bag 4, Ganeshkhind, Pune 411~007, India}\\
{\small \it $^2$ RGGR, Facult\'{e} des Sciences }\\
{\small \it Campus Plaine, Universit\'{e} Libre de Bruxelles}\\
{\small \it Boulevard du Triomphe, CP 231}\\
{\small \it 1050 Bruxelles, Belgium}\\
{\small \it $^3$ Institutes Internationaux de Chimie et Physique Solvay}
}
\date{}
\maketitle   \thispagestyle{empty}  \vspace*{1truecm}

\begin{abstract}

The amplification of a light beam due to intervening gravitational waves 
is studied. The previous Jordan frame result according to which the 
amplification is many orders of magnitude larger in scalar--tensor 
gravity than in general relativity does not hold in the 
Einstein conformal frame. Lensing by gravitational waves is discussed in 
relation to the ongoing and proposed {\em VLBI} observations aimed at 
detecting the scintillation effect. 
\end{abstract} \vspace*{1truecm} 
\begin{center}
To appear in {\em Astronomy \& Astrophysics} 
\end{center}
\end{titlepage}   \clearpage

\section{Introduction}

Among the proposed theories of gravity, a special position is occupied by
scalar--tensor theories, which currently are the subject of great 
interest because they exhibit 
features that resemble those of string theories (Green, Schwarz \& Witten
1987). First of all, a fundamental scalar field $\phi$ appears in 
scalar--tensor theories in addition to
the metric tensor $g_{\mu\nu}$, and massless scalar fields 
coupled to gravity are an essential feature of string and supergravity 
theories. Second, the gravitational part of the scalar--tensor action, 
\be
S_g=\frac{1}{16\pi}\int d^4x \sqrt{-g} \left[ \phi \, R -\frac{\omega( \phi 
)}{\phi} \nabla_{\alpha}\phi \nabla^{\alpha}\phi \right]\; ,
\ee
exhibits a conformal invariance that mimics the conformal invariance of
string theories at high energies (Cho 1992, 1994; Damour
\& Esposito--Farese
1992;  Turner 1993; Kolitch \& Eardley 1995; Brans 1997).
Further motivation for the study of scalar--tensor theories comes from the
extended and hyperextended inflationary scenarios of the early universe (La
\& Steinhardt 1989; Steinhardt \& Accetta 1990; Kolb, Salopek \& Turner
1990; Liddle \& Wands 1992; Crittenden \& Steinhardt 1992;  Steinhardt
1993; Laycock \& Liddle 1994). Given that the classical tests of gravity
in the Solar System (e.g. Will 1993) tell us that gravity is very close to
Einstein gravity today\footnote{It is possible that gravity 
was described by a scalar--tensor theory early in the history of the 
universe, and converged to general relativity in the era of 
matter domination (Damour \& Nordvedt 1993a,b). If this is the 
case, the possibility of testing relativistic gravity at high redshifts 
is even more attractive.}, any experiment that allows one to 
discriminate between general relativity and an alternative theory of 
gravity with present technology is important. An astronomical 
effect with such a 
potentiality was pointed out recently (Faraoni 1996); by studying the 
propagation of a light beam through gravitational waves, it was shown 
that the 
time--dependent amplification induced in the beam is a first order effect 
in the gravitational wave amplitudes, in scalar--tensor theories. This
is an improvement of several orders of magnitude over the case of general 
relativity, in 
which the effect is only of second order (Bertotti 1971). The study of 
this effect is particularly relevant for  
the {\em VLBI} observations presently carried out on the radio 
source 2022+171 (Pogrebenko et al. 1994a,b, 1996) or 
proposed by Labeyrie (1993) (see also Bracco 1997) in 
order to detect the scintillation induced by gravitational waves. In 
recent years, many theoreticians have devoted their attention to the 
action of gravitational waves as lenses (Braginsky et al. 1990; 
Faraoni 1992a,b, 1993, 1996, 1997; Fakir 1993, 1994a,b, 1995, 
1997; Durrer 1994; Marleau \& Starkman 1996; Kaiser \& Jaffe 1997; 
Gwinn et al. 1997), 
or as perturbations of conventional gravitational lenses (McBreen \& 
Metcalfe 1988; Allen 1989, 1990; Kovner 1990; Frieman, Harari \& 
Surpi 1994; Bar--Kana 1996).

The analysis of the amplification effect in (Faraoni 1996) was 
performed in the Jordan conformal frame and was based on 
the propagation equations for  the optical scalars\footnote{The metric 
signature is --~+~+~+, the speed of light and Newton's 
constant are set equal to unity, a colon and a semicolon denote, 
respectively, ordinary and covariant differentiation, $\nabla_{\mu}$ is the 
covariant derivative operator. Round and square brackets around indices 
denote, respectively, symmetrization and antisymmetrization. The Ricci 
tensor is given in terms of the Christoffel 
symbols $\Gamma_{\alpha\beta}^{\delta}$ by 
$ R_{\mu\rho}=
\Gamma^{\nu}_{\mu\rho ,\nu}-\Gamma^{\nu}_{\nu\rho ,\mu}+
\Gamma^{\alpha}_{\mu\rho}\Gamma^{\nu}_{\alpha\nu}-
\Gamma^{\alpha}_{\nu\rho}\Gamma^{\nu}_{\alpha\mu}$, and
$\Box \equiv g^{\mu\nu}\nabla_{\mu}\nabla_{\nu}$. A tilde denotes 
quantities defined in the Einstein conformal frame.} (Sachs 1961)
\begin{eqnarray}    
&& \theta=\frac{1}{2} \,
{k^{\alpha}}_{;\alpha} \; ,  \\
&&  \sigma=\frac{1}{\sqrt{2}}
\left[
k_{( \alpha;\beta)}k^{\alpha;\beta}-\frac{1}{2} \left(
{k^{\alpha}}_{;\alpha}
\right)^2 \right]^{1/2} \; , \\
&& \omega =\frac{1}{\sqrt{2}} \, \left[ k_{[\alpha;\beta]} \, 
k^{\alpha;\beta} \right]^{1/2} \: ,
\end{eqnarray} 
of a congruence of null rays with tangent vector field $k^{\mu}$. In 
Einstein gravity, the study of the Raychaudhuri equation
\be   \label{Raychaudhuri}
\frac{d\theta}{d\lambda}=-\,\theta^2-|\sigma|^2+\omega^2- \frac{1}{2} \,
R_{\mu\nu}k^{\mu}k^{\nu} 
\ee
(where $\lambda$ is an affine parameter along the null geodesics) in the 
metric 
\be   \label{JFdecomposition}
g_{\mu\nu}=\eta_{\mu\nu}+h_{\mu\nu}
\ee
(the perturbations 
$h_{\mu\nu}$, with $|h_{\mu\nu}| \ll 1$, describe gravitational waves)
leads, to first order in the waves amplitude 
$h$, to a vanishing Ricci tensor, and to the solution 
$\theta_{GR}=$O($h^2$) for the expansion $\theta $ of the congruence 
(Bertotti 1971). 
This quantity describes the amplification of the beam in the geometric 
optics approximation, since the photon number is conserved 
(Schneider, Ehlers \& Falco 1992). In scalar--tensor theories 
formulated in the Jordan conformal frame, one also has a gravitational 
scalar field $\phi=\phi_0 +\varphi$, where $\phi_0$ is constant and O$ 
\left( \varphi/\phi_0 \right) =$O$( h) $. This leads to a nonvanishing 
term $R_{\alpha\beta}k^{\alpha} k^{\beta}$ on the right hand side of  
Eq.~(\ref{Raychaudhuri}), which corresponds to a form of matter (scalar 
waves) in the beam; the first order amplification $\theta_{JF}=$O$(h)$ 
arises as a consequence (Faraoni 1996). However,  
the expression $-R_{\mu\nu} k^{\mu} k^{\nu} $ oscillates with the 
frequency of $\varphi$, and this is a disturbing signal of the 
violation of the weak energy condition. The fact that 
$-R_{\mu\nu} k^{\mu} k^{\nu} $ is always negative whenever the energy 
conditions are satisfied, is essential in the proof of the singularity 
theorems (Wald 1984), hence the old adagio ``matter always focuses''. 
The anomaly is in fact due to the violation of the weak energy 
condition in the Jordan frame, as will be explained in Sec.~2.
Focusing of null geodesics is caused by the energy of the waves, and the 
anomalous dependence of the energy (linear in the second derivatives of 
the field, instead of quadratic in its first derivatives) in the Jordan 
frame version of scalar--tensor theories is reflected in the lensing effect.

It could be objected that a time--average gets rid of the offending 
first order expression; however, the problem is not solved. 
One can consider gravitational waves of astrophysical interest with 
relatively long periods (e.g. waves from $\mu$--Sco, with period $3\cdot 
10^5$~s), for which the weak energy condition is violated on physically 
significant time scales.

There have been many debates in the literature on the issue of the 
conformal frame, which is still the subject of controversy. We do not 
repeat here these discussions but, rather, we refer the reader to 
(Magnano \& Sokolowski 1994, and references therein). For our purposes, 
it is sufficient to remember that, in the Jordan frame formulation of 
scalar--tensor theories, the kinetic energy term 
for the scalar field in the Jordan action has indefinite sign, the 
system decays into a lower energy state {\em ad infinitum}, and it is 
unstable against small fluctuations. On the contrary, the reformulation 
of the scalar--tensor theory in the Einstein conformal frame exhibits a 
positive definite, canonical kinetic term for the Brans--Dicke--like 
scalar, and the theory has the desired stability 
property (Magnano \& Sokolowski 1994, and references therein). The 
present paper rephrases these arguments in terms of the weak energy 
condition~--~The reader should be warned that many current papers and most 
textbooks still present only the Jordan frame version of scalar--tensor 
theories.

The metric $\tilde{g}_{\mu\nu}$ in the Einstein frame is related to the
Jordan frame metric $ g_{\mu\nu}$ by 
the conformal transformation
\be          \label{CT}
g_{\mu\nu} \longrightarrow \tilde{g}_{\mu\nu}=\Omega^2 g_{\mu\nu}
\;\;\;\;\; , \;\;\;\;\; \Omega=\sqrt{\phi } \; ,
\ee
and the scalar fields in the two frames are related by the redefinition
\be          \label{SFredefinition}
\phi \longrightarrow \tilde{\phi}=\int \frac{ \left( 2\omega+3 
\right)^{1/2}}{\phi}\, d\phi \; ,
\ee
where $\omega >-3/2$. The necessity of the conformal transformation and 
arguments supporting the Einstein frame formulation were first advocated in 
Kaluza--Klein and Brans--Dicke theories (Sokolowski \& Carr 1986; 
Bombelli et al. 1987; Sokolowski \& Golda 1987; Sokolowski 1989a,b; Cho
1990, 1994) and later generalized to scalar--tensor theories 
(Cho 1997) and non--linear gravity theories with gravitational part of 
the Lagrangian ${\cal L}=f( \phi, R ) $ (Magnano \& Sokolowski 1994, and 
references therein). 
It is important to reanalyse the calculations of (Faraoni 1996) in the 
Einstein conformal frame and to compute the magnitude of the 
amplification effect. The new calculation is presented in Sec.~2; photons 
propagating through a cosmological background of scalar--tensor 
gravitational waves are considered in Sec.~3, while Sec.~4 contains the 
conclusions.

\section{Einstein frame vs Jordan frame}

Since the Maxwell equations in four dimensions are conformally invariant, 
photons follow null geodesics also in the Einstein frame, as expected 
in the geometric optics approximation. We begin by decomposing the 
Einstein frame metric and scalar field as follows: 
\setcounter{equation}{0} 
\be  \label{EFM}
\tilde{g}_{\mu\nu}=\eta_{\mu\nu}+\tilde{h}_{\mu\nu} \; , 
\ee 
\be        \label{EFSF}
\tilde{\phi}=\tilde{\phi}_0+\tilde{\varphi} \; , 
\ee 
where
$\tilde{\phi}_0=$constant and $\tilde{h}_{\mu\nu}$, 
$ \tilde{\varphi} $
describe, respectively, tensor and scalar gravitational waves, with
$|\tilde{h}_{\mu\nu}|$, $ \left| \tilde{\varphi}/\tilde{\phi_0} \right| 
\ll 1 $.  The linearized field equations are 
\be \label{EFFE1}
\tilde{\Box} \bar{\tilde{h}}_{\mu\nu}=0 \; , 
\ee 
\be  \label{EFFE2} 
\tilde{\Box} \tilde{\varphi}=0 \; , 
\ee 
where $\bar{\tilde{h}}_{\mu\nu} \equiv \tilde{h}_{\mu\nu}-\eta_{\mu\nu} 
\tilde{h}^{\alpha}_{\alpha}/2$. The solutions of 
Eq.~(\ref{EFFE2}) are expressed as Fourier integrals of plane waves,
\be    \label{planewaves}
\tilde{\varphi}=\tilde{\varphi}_0 \cos \left( p_{\alpha} x^{\alpha} 
\right) \; ,
\ee
where $\tilde{\varphi}_0$ is a constant and $\eta_{\mu\nu} 
p^{\mu}p^{\nu}=0$. The stress--energy tensor of the scalar field assumes 
the canonical form 
\be    \label{Tmunu}
\tilde{T}_{\mu\nu}[ \tilde{\varphi} ]= \partial_{\mu} 
\tilde{\varphi} \partial_{\nu} \tilde{\varphi}-\frac{1}{2} \eta_{\mu\nu} 
\eta^{\alpha\beta} \partial_{\alpha} \tilde{\varphi}
\partial_{\beta} \tilde{\varphi} \; .
\ee
The term $\tilde{R}_{\mu\nu} 
k^{\mu} k^{\nu}$ on the right hand side of the Raychaudhuri 
equation (\ref{Raychaudhuri}), which is responsible for the first order 
amplification effect in the Jordan frame, is nonzero also in the Einstein 
frame, but it is now of second order. Equations~(\ref{Tmunu}) and 
(\ref{planewaves}) yield\footnote{The order of magnitude of the 
perturbations is the same in both conformal frames (see the Appendix).} 
\be
\tilde{R}_{\mu\nu} k^{\mu} k^{\nu}= 4\pi \left[ p_{\alpha}k^{\alpha} 
 \tilde{\varphi}_0  \sin ( p_{\alpha}x^{\alpha} )    \right]^2 
+\tilde{T}_{\mu\nu}^{(eff)} [ \tilde{h}_{\alpha\beta} ] k^{\mu} k^{\nu} \; , 
\ee
which is of second order and positive definite; 
$\tilde{T}_{\mu\nu}^{(eff)} [ \tilde{h}_{\alpha\beta} ]$ is the Isaacson 
effective stress--energy tensor of the tensor modes 
$\tilde{h}_{\alpha\beta}  $. Following the reasoning of (Bertotti 1971; 
Faraoni 1996), which we do not repeat here, it is straightforward to 
conclude that the amplification of the beam in the Einstein frame 
is of second order, $\theta_{EF}=$O$( \tilde{h}^2)$, contrarily to the 
case of the Jordan frame. Since $| h_{\mu\nu} | \ll 1$, this changes the 
amplification by many orders of magnitude.

Why is there such a difference between the Jordan and the Einstein 
frame~? This is due to the different orders of magnitude 
of the term $R_{\mu\nu} k^{\mu} k^{\nu}$ in the Raychaudhuri equation. 
The Ricci tensor changes under the conformal transformation (\ref{CT}) 
according to (Wald 1984) 
\begin{eqnarray}   
& & \tilde{R}_{\alpha\beta}=R_{\alpha\beta}-2\nabla_{\alpha}\nabla_{\beta}( 
\ln \Omega )-g_{\alpha\beta}g^{\rho\sigma} \nabla_{\rho}\nabla_{\sigma} 
( \ln \Omega ) +2 \nabla_{\alpha} ( \ln \Omega ) \nabla_{\beta} ( \ln 
\Omega ) \nonumber \\ 
& & -2 g_{\alpha\beta} g^{\rho\sigma}\nabla_{\rho} ( \ln \Omega ) 
\nabla_{\sigma} ( \ln \Omega ) \; ; \label{EFRicci}
\end{eqnarray}
the harmonic expansion of the Jordan frame scalar 
\be    \label{JFSF}
\phi=\phi_0+\varphi_0 \cos ( l_{\alpha}x^{\alpha} ) 
\ee 
yields the first order term $R_{\mu\nu}=\partial_{\mu}\partial_{\nu} 
\varphi/\phi_0  $ (Eq.~(6) of (Faraoni 1996)).
This expression, which is the first term on the right hand side of 
Eq.~(\ref{EFRicci}), is exactly cancelled by the first order contribution 
to the next term $-2\nabla_{\alpha}\nabla_{\beta} 
(  \ln \Omega )=-\partial_{\alpha}\partial_{\beta} \varphi /  \phi_0  
+ $O($h^2$); what is left on the right hand side of Eq.~(\ref{EFRicci})  
is only of second order.

The expansion (\ref{JFSF}) is, of course, consistent
with Eqs.~(\ref{EFSF}), (\ref{planewaves}); in fact, from 
Eqs.~(\ref{JFSF}), (\ref{SFredefinition}) it follows that 
\be    \label{consistency}
\tilde{\phi}=\frac{\left( 2\omega_0+3 \right)^{1/2}}{\phi_0} \, \varphi 
+C+\mbox{O}(h^2) \; ,
\ee
where $\omega_0=\omega ( \phi_0 )$ and $C$ is a
integration constant.  To first order, Eq.~(\ref{consistency}) is nothing 
but Eq.~(\ref{planewaves}) where
\be 
\tilde{\phi}_0=C \; ,
\ee
\be
\tilde{\varphi}_0 = \frac{ \left( 2\omega_0+3 \right)^{1/2}}{\phi_0} 
\, \varphi_0 \; ,
\ee
\be
p^{\alpha}=l^{\alpha} \; .
\ee
The origin of the problem in the Jordan frame is the non--canonical form 
of the stress--energy tensor of the scalar field; the $T_{\mu\nu}[ \phi ] 
$ of the Brans--Dicke scalar in the Jordan frame violates the weak energy 
condition, and its structure is also responsible for the first order 
amplification. For simplicity, we restrict our treatment to Brans--Dicke 
theory, in which $\omega$ is constant, with vanishing cosmological 
constant. Gravitational waves in the Jordan frame are then described by the 
metric and scalar field perturbations $h_{\mu\nu}$ and $\varphi$ in 
Eqs.~(\ref{JFdecomposition}), (\ref{JFSF}). The field equations yield the 
linearized equations (Will 1993) 
\be
\Box_{\eta} \varphi=0 \; ,
\ee
\be
R_{\mu\nu}-\frac{1}{2} \eta_{\mu\nu}R= \frac{1}{\phi_0} 
\partial_{\mu}\partial_{\nu} \varphi \; .
\ee
By using the decomposition of $\varphi$ in plane waves one has that, for 
each plane monochromatic wave, 
\be
T_{\mu\nu} [ \varphi ] \xi^{\mu} \xi^{\nu} = -\left( k_{\mu} \xi^{\mu} 
\right)^2 \frac{\varphi}{\phi_0} 
\ee
for any timelike vector $\xi^{\mu}$. Since $\varphi$ is an oscillating 
quantity, the sign of the energy density measured by an observer with 
four--velocity $\xi^{\mu}$ changes with the frequency of $\varphi$, 
violating the weak energy condition. By contrast, the Einstein frame 
stress--energy tensor is the sum of the canonical tensor for a scalar 
field, plus the effective Isaacson tensor for spin~2 waves: 
\be
\tilde{T}_{\mu\nu} [ \tilde{\varphi} ] =
 \tilde{\nabla}_{\mu} \tilde{\varphi} \tilde{\nabla}_{\nu} 
\tilde{\varphi} 
-\frac{1}{2} \tilde{g}_{\mu\nu} \tilde{\nabla}^{\alpha} \tilde{\varphi} 
\tilde{\nabla}_{\alpha} \tilde{\varphi}
+\tilde{T}_{\mu\nu}^{(eff)} [ \tilde{h}_{\alpha\beta} ] =
{\mbox O}( h^2) \; .
\ee
One obtains, to the lowest order in the Einstein frame, 
\be
\tilde{T}_{\mu\nu} \xi^{\mu} \xi^{\nu} =
\left[ \xi^{\alpha} p_{\alpha} \varphi_0 \sin ( p_{\beta} x^{\beta} ) 
\right]^2 
+\tilde{T}_{\mu\nu}^{(eff)} [ \tilde{h}_{\alpha\beta} ] \xi^{\mu} \xi^{\nu}
\geq 0 \; . 
\ee 
Besides violating the weak energy condition,  the non--canonical form of 
$T_{\mu\nu} [ \varphi ]$ in the Jordan frame 
is also responsible for the order of magnitude of the term 
$R_{\mu\nu} k^{\mu} k^{\nu}$ in the Raychaudhuri equation;  
$T_{\mu\nu} [ \varphi ] $ is not a quadratic form in the scalar field 
derivatives but contains a term that depends linearly from the 
second derivatives of $\varphi$~--~this is the only term that survives for 
linearized waves. By contrast, the Einstein frame $\tilde{T}_{\mu\nu} [ 
\tilde{\varphi}] $ complex (associated to the usual energy functional) is 
quadratic in the scalar field derivatives, and hence it is positive 
definite.

\section{The gravitational wave background}

The propagation of light through the cosmological gravitational wave 
background (Matzner 1968) has been studied in Einstein gravity in  order 
to ascertain 
whether the deflection and frequency shift of the photons (which are 
of first order in the gravitational wave amplitudes, and therefore small) 
cumulate with the travelled distance $D$. Since $D$ can be a cosmological 
distance, this secular or ``$D$--effect'', if present, would 
significantly enhance the deflections and frequency shifts, and it was 
considered also in relation with redshift anomalies and periodicities in 
galaxy groups and clusters (Rees 1971; Dautcourt 1974), and with proper 
motions of quasars (Gwinn et al. 1997).
Naively, one would expect that, if a photon undergoes $N$ scatterings in a 
background of random gravitational waves, the deflections add
stochastically, resulting in a $D$--effect proportional to $\sqrt{N}$  
(Winterberg 1968; Marleau \& Starkman 1996). This is not 
the case, due to the equality between the speed of 
the propagating signals and that of the random inhomogeneities of the 
medium (Zipoy 1966; Zipoy \& Bertotti 1968; Dautcourt 1974; Bertotti \& 
Catenacci 1975; Linder 1986, 1988; Braginsky et al. 1990; Kaiser \&  
Jaffe 1997). Is the $D$--effect present 
in a stochastic background of scalar--tensor gravitational waves~? This 
question is non--trivial because random inhomogeneities due to fields of 
different spin produce different results for the rms deflection, and 
spin~0 waves go hand in hand with spin~2 modes in scalar--tensor gravity.

The solution to this problem is contained in Linder's (1986) paper; 
although he did not explicitely consider alternative theories of gravity, 
he studied light propagation through a random medium with inhomogeneities 
due to fields of spin~0, 1 or 2, which are allowed to propagate at any 
speed less than, or equal to, the speed of light. Adapting Linder's 
(1986) result to the case of scalar modes propagating at the speed of 
light, one obtains that a photon whose unperturbed path is 
parallel to the $z$--axis, experiences the rms deflection 
\setcounter{equation}{0}
\be
\left( \delta k^x \right)_{rms}=\left( \delta k^y \right)_{rms}=
\left( \frac{\tilde{\varphi}}{\tilde{\phi}_0} \right)_{rms} \left[ \ln 
\left( \frac{2\pi D}{\lambda_{gw}} \right) \right]^{1/2} \; ,
\ee
\be 
\left( \delta k^z \right)_{rms}=0 \; .
\ee
The same dependence was obtained in a recent paper by Kaiser \& Jaffe 
(1997). Albeit qualitatively different from Einstein gravity, the dependence 
of $\left( \delta k^{\mu} \right)_{rms} $ from the distance $D$ is hardly 
significant: to give an idea of the orders of magnitude involved, we 
consider a gravitational wavelength $\lambda_{gw}=5$~cm 
(approximately corresponding to the 1~K cosmic gravitational wave 
background (Matzner 1968)) and a cosmological distance $D =500$~Mpc, which 
give
\be
\left( \delta k \right)_{rms}\simeq 7.92 
\left( \frac{\tilde{\varphi}}{\tilde{\phi}_0} \right)_{rms} \; ,
\ee
an enhancement of less than one order of magnitude with respect to general 
relativity. Again, lensing by gravitational waves in scalar--tensor 
gravity is not much more efficient than in Einstein gravity.

\section{Conclusions}

From the theoretical point of view, our analysis is relevant to the issue
of the conformal frame in scalar--tensor theories of gravity. The 
violation of the weak energy condition in the Jordan frame was shown in 
~Sec.~2. From the
point of view of the applications of the theory, we have given a negative
answer to the problem of whether, in scalar--tensor theories, the gravity
wave--induced amplification of a light beam is enhanced by many orders of
magnitude in comparison to Einstein gravity. If this was true, a door
would be open for discriminating between general relativity and
scalar--tensor theories using astronomical observations and present
technology. Our study is relevant for the ongoing {\em VLBI} observations
of the radio source 2022+171 (Pogrebenko et al. 1994a,b, 1996) aimed at
detecting gravity wave--induced scintillation effects, and in view of the
observations proposed by Labeyrie (1993) (see also Bracco 1997).
Unfortunately, when the amplification of a light beam due to gravitational
waves is computed in the Einstein conformal frame, to which the
observations must be referred, the effect is not much larger in
scalar--tensor gravity than it is in general relativity. This leaves
little hope for an easy detection of the scintillation effect, and for the
determination of the correct theory of gravity using astronomy. This
rather pessimistic conclusion is based on the assumption that scalar and
tensor modes have the same amplitude, O$(\tilde{\varphi}/\tilde{\phi}_0
)=$O$( \tilde{ h}_{\mu\nu})$; perhaps this assumption is relaxed to a
certain extent if scalar modes are emitted at a significantly larger rate
than tensor modes in processes of astrophysical relevance. For example,
gravitational collapse with spherical symmetry produces spin~0, but not
spin~2, waves.

Taking a broader point of view, it would be premature to conclude that the 
amplification induced by gravitational waves (in general relativity or in 
its scalar--tensor competitors) is impossibile to detect with present 
technology. In fact, the optical scalars formalism used in our 
calculation breaks down when the gravitational lens exhibits caustics and 
critical lines, which separate regions corresponding to different numbers 
of images of the light source. In this situation, high amplification events 
occur if the light source crosses a caustic (Schneider, Ehlers \& 
Falco 1992). In (Faraoni 1997), it was shown that the optical scalars 
formalism does not tell the whole story: the actual amplification is of 
order 
\setcounter{equation}{0}
\be
{\cal A} \approx \left( \frac{hD}{\lambda_{gw}} \right)^2 \; ,
\ee
where $\lambda_{gw} $ is the gravitational wavelength, and $D$ is the 
distance between the observer and the light source. Large values of the 
ratio $D/\lambda_{gw}$ can balance small values of $h$ and a 
non--negligible amplification is still possible. A detailed study of 
realistic 
gravitational waveforms within the formalism developed in (Faraoni 
1992a,b, 1997) and a feasibility study of the {\em VLBI} detection 
of scintillation effects induced by gravitational waves will be the
subject of a 
future publication. The conclusion of the present paper is that 
scalar--tensor gravity does not fare much better than  general relativity 
in inducing this kind of effects.

\section*{Acknowledgments}

We are grateful to C. Bracco, P. Teyssandier and A. Labeyrie for stimulating 
discussions. VF would like to acknowledge also B. Bertotti for introducing 
him to the topic of lensing by gravitational waves.

This work was partially supported by EEC grants numbers PSS*~0992 and
CT1*--CT94--0004, and by OLAM, Fondation pour la Recherche Fondamentale,
Brussels.

\clearpage
\section*{Appendix}

Insight into the nature of scalar--tensor waves in the Einstein 
frame is obtained by combining Eq.~(\ref{CT}) and the metric
decompositions (\ref{JFdecomposition}), (\ref{EFM}) to obtain
\def\theequation{A.\arabic{equation}}\setcounter{equation}{0}
\be   \label{mixmodes}
\tilde{h}_{\mu\nu}=h_{\mu\nu}+\frac{\varphi}{\phi_0} \eta_{\mu\nu} + 
{\mbox O}(h^2) \; .
\ee
According to Eq.~(\ref{mixmodes}), the Einstein frame gravitational 
waves are a mixture of spin~2 ($h_{\mu\nu}$) and spin~0 ($\eta_{\mu\nu} 
\varphi/\phi_0 $) modes in the language of the Jordan frame. Moreover, 
the metric perturbations have the same order of magnitude in the two 
conformal frames: 
\be
{\mbox O}( \tilde{h}_{\mu\nu} )={\mbox O}( h_{\mu\nu}) \; ,
\ee
\be
{\mbox O} \left( \frac{\tilde{\varphi}}{\tilde{\phi_0}} \right)=
{\mbox O} \left( \frac{\varphi}{\phi_0} \right)
\ee
(where the last equality follows from Eq.~(\ref{consistency}). It is this  
property that allows a meaningful comparison of the amplification effect 
in the two conformal frames. 

\clearpage
{ \small 
\section*{References}

\noindent Accetta, F.S., Steinhardt, P.J., 1990, Phys. Rev. Lett. 64,
2740\\ 
Allen, B., 1989, Phys. Rev. Lett. 63, 2017\\
Allen, B., 1989, Gen. Rel. Grav. 22, 1447\\ 
Bar--kana, R., 1996, Phys. Rev. D 54, 7138\\ 
Bertotti, B., 1971, in Sachs, R.K. (ed.)  General Relativity and 
Cosmology, XLII Course of the Varenna Summer School, Academic Press, New
York, p.~347\\ 
Bertotti, B., Catenacci, R., 1975, Gen. Rel. Grav. 6, 329\\ 
Bombelli, L., Koul, R.K., Kunstatter, G., Lee, J., Sorkin, R.D., 1987,
Nucl. Phys. B 289, 735\\ 
Bracco, C., 1997, A \& A 321, 985\\ 
Brans, C.H., 1997, preprint gr-qc/9705069\\ 
Braginsky, V.B., Kardashev, N.S., Polnarev, A.G., Novikov, I.D., 1990,
Nuovo Cimento 105B, 1141\\ 
Cho, Y.M., 1990, Phys. Rev. D 41, 2462\\
Cho, Y.M., 1992, Phys. Rev. Lett. 68, 3133\\ 
Cho, Y.M., 1994, in Sato, H. (ed.) Evolution of the Universe and its
Observational Quest, Proceedings, Yamada, Japan 1993, Universal Academy
Press, Tokyo, p.~99\\ 
Cho, Y.M., 1997, Class. Quant. Grav. 14, 2963\\ 
Crittenden, R., Steinhardt, P.J., 1992, Phys. Lett. B 293, 32\\ 
Damour, T., Esposito--Farese, G., 1992, Class. Quant. Grav. 9, 2093\\
Damour, T., Nordvedt, K., 1993a, Phys. Rev. D 48, 3436\\ 
Damour, T., Nordvedt, K., 1993b, Phys. Rev. Lett. 70, 2217\\ 
Dautcourt, G., 1974, in Longair M.S. (ed.) Confrontation of
Cosmological Theories with Observation, Proc. IAU Symp.~63, Reidel,
Dordrecht\\ 
Durrer, R., 1994, Phys. Rev. Lett. 72, 3301 \\
Fakir, R., 1993, ApJ 418, 202\\ 
Fakir, R., 1994a, Phys. Rev. D 50, 3795 \\ 
Fakir, R., 1994b, ApJ 426, 74 \\ 
Fakir, R., 1995, preprint astro--ph/9507112\\ 
Fakir, R., 1997, Int. J. Mod. Phys. D 6, 49\\ 
Faraoni, V., 1992a, in Kayser, R., Schramm, T., Nieser, L. (eds.)
Gravitational Lenses, Proceedings,  Hamburg 1991,  Springer--Verlag, 
Berlin\\
Faraoni, V., 1992b, ApJ 398, 425 \\ 
Faraoni, V., 1996, Astrophys. Lett. Comm. 35, 305\\ 
Faraoni, V., 1997, Int. J. Mod. Phys. D, in press (preprint IUCAA--51/97, 
astro--ph/9707236) \\ 
Frieman, J.A., Harari, D.D., Surpi, G.C., 1994, Phys. Rev. D 50, 4895\\ 
Green, B., Schwarz, J.M., Witten, E., 1987, Superstring Theory, 
Cambridge University Press, Cambridge\\ 
Gwinn, C.R., Marshall Eubanks, T., Birkinshaw, M., Matsakis, D.N., 
1997, ApJ 485, 87\\
Kaiser, N., Jaffe, A., 1997, ApJ 484, 545\\ 
Kolb, E.W., Salopek, D., Turner, M.S., 1990, Phys. Rev. D 42, 3925\\ 
Kolitch, S.J., Eardley, D.M. 1995, Ann. Phys. (NY) 241, 128\\ 
Kovner, I., 1990, ApJ 351, 114\\ 
La, D., Steinhardt, P.J., 1989, Phys.  Rev. Lett. 62, 376\\ 
Labeyrie, A., 1993, A \& A 268, 823\\ 
Laycock, A.M., Liddle, A.R., 1994, Phys. Rev. D, 49, 1827\\ 
Liddle, A.R., Wands, D., 1992, Phys. Rev. D 45, 2665\\ 
Linder, E.V., 1986, Phys. Rev. D 34, 1759\\ 
Linder, E.V., 1988, ApJ 328, 77\\ 
Magnano, G., Sokolowski, L.M. 1994, Phys. Rev. D 50, 5039\\ 
Marleau, F.R., Starkman, G.D. 1996, preprint astro--ph/9605066\\ 
Mc Breen, B., Metcalfe, L. 1988, Nat 332, 234\\ 
Matzner, R.A., 1969, ApJ 154, 1085\\ 
Pogrebenko, S. et al., 1994a, in Kus, A.J., Schilizzi, R.T., 
Borkowski,K.M., Gurvits, L.I. (eds.),
Proc. 2nd EVN/JIVE Symposium, Torun, Poland 1994, Torun Radio Astronomy
Observatory, Torun Poland, p.~33\\ 
Pogrebenko et al., 1994b,
Abstracts XXIInd GA IAU Meeting, Twin Press, Netherlands, p.~105\\ 
Pogrebenko et al. 1996, in Van
Paradijs, J., Van den 
Heuvel, E.P.J., Kuulkers, E. (eds.) Compact Stars 
in Binaries, Proc. IAU Symp. 165, The Hague, Netherlands 1994, Kluwer,
Dordrecht, p.~546\\ 
Rees, M.J. 1971, MNRAS 154, 187\\ 
Sachs, R.K., 1961, Proc. Roy. Soc. Lond. A 264, 309\\  
Schneider, P., Ehlers, J., Falco, E.E., 1992, Gravitational Lenses, 
Springer, Berlin\\ 
Sokolowski, L., 1989a, Class. Quant. Grav. 6, 59\\ 
Sokolowski, L., 1989b, Class. Quant. Grav. 6, 2045\\ 
Steinhardt, P.J., 1993, Class. Quant. Grav. 10, S33\\
Steinhardt, P.J., Accetta, F.S., 1990, Phys. Rev. Lett. 64, 2740\\ 
Turner, M.S., 1993, in Harvey, J., Polchinski, J. (eds.) Recent
Directions in Particle Theory -- From Superstrings and Black Holes to the
Standard Model, Proc. Theoretical Advanced Study Institute in Elementary
Particle Physics, Boulder, Colorado 1992, World Scientific, Singapore\\ 
Wald, R.M., 1984, General Relativity, The University of Chicago Press, 
Chicago\\ 
Will, C.M., 1993, Theory and Experiment in Gravitational Physics,
Cambridge University Press, Cambridge \\ 
Winterberg, F., 1968, Nuovo Cimento 53B, 1096 \\ 
Zipoy, D.M. 1966, Phys. Rev. 142, 825 \\
Zipoy, D.M., Bertotti, B. 1968, Nuovo Cimento 56B, 195

}    
\end{document}